\documentclass[twocolumn,english,pra,aps]{revtex4-1}

\usepackage{ae,aecompl}
\usepackage{helvet}
\usepackage[T1]{fontenc}
\usepackage[latin9]{inputenc}
\usepackage{babel}
\usepackage{amsmath}
\usepackage{amssymb}
\usepackage{lipsum}
\usepackage{graphicx}
\usepackage{esint}
\usepackage{hyperref}

\makeatletter
\@ifundefined{textcolor}{}
{%
 \definecolor{BLACK}{gray}{0}
 \definecolor{WHITE}{gray}{1}
 \definecolor{RED}{rgb}{1,0,0}
 \definecolor{GREEN}{rgb}{0,1,0}
 \definecolor{BLUE}{rgb}{0,0,1}
 \definecolor{CYAN}{cmyk}{1,0,0,0}
 \definecolor{MAGENTA}{cmyk}{0,1,0,0}
 \definecolor{YELLOW}{cmyk}{0,0,1,0}
}

\makeatother

\begin{document}

\title{Generalized Tunneling Model for TLS in amorphous materials and its
predictions for their dephasing and the noise in superconducting microresonators}

\author{Lara Faoro$^{1,2}$ and Lev B. Ioffe$^{1,2}$}

\affiliation{$^{1}$ Laboratoire de Physique Theorique et Hautes Energies, CNRS
UMR 7589, Universites Paris 6 et 7, 4 place Jussieu, 75252 Paris,
Cedex 05, France}

\affiliation{$^{2}$ Department of Physics and Astronomy,~Rutgers The State University
of New Jersey, 136 Frelinghuysen Rd, Piscataway,~08854 New Jersey,
USA }

\begin{abstract}
We formulate the generalized tunneling model for two level systems
in insulators that takes into account the interaction between them
and a slow power law dependence of their density of states. We show
that the predictions of this model are in a perfect agreement with
the recent studies of the noise in high quality superconducting resonators.
The predictions also agree with the temperature dependence of the TLS dephasing rates
observed in phase qubits. Both observations are impossible to
explain in the framework of the standard tunneling model of TLS. We
discuss the origin of the universal dimensionless parameter that controls
the interaction between TLS in glasses and show that it is consistent
with the assumptions of the model. 
\end{abstract}

\pacs{85.25.Cp, 03.65.Yz,73.23.-b}

\maketitle

\section{Introduction}

Thin film high quality superconducting resonators are important for
a number of applications, ranging from quantum computation to submillimeter
and far-infrared astronomy \cite{Zmuidzinas2012}. The performance
of these devices has improved dramatically over the past decades and
resonator quality factors above $10^{6}$ are now routinely fabricated
using single-layer superconductors deposited on high quality low-loss
crystalline substrate. Achieving resonators with high quality factors requires
minimization of all potential sources of dissipation and noise. 

The major source of dissipation and noise in the resonators is Two
Level Systems (TLS) located in the amorphous dielectrics. 
The presence and importance of TLS was proven by the measurements of the resonators frequency
shifts as a function of the temperature \cite{Gao2008}. These experiments
have shown unambiguously that even in the devices that do no use a
deposited dielectric and consist only of a patterned superconducting
film on high-quality crystalline substrate, a thin, TLS-hosting layer
is present on the surface of the device. In particular, TLS in the
thin amorphous surface layer of the microresonators are responsible
for the noise in the resonator frequency that is the subject of this
paper. This noise has been carefully characterized in the last few
years. The early works reported unusual behavior of the noise spectral
density, $S\sim f^{-1/2}$ \cite{Gao2007,Gao2008,Kumar2008,Barends2009,Barends2010} but all
recent works \cite{Burnett2013,Burnett2014,Neill2013,Murch2012} agree
on a more conventional $S\sim1/f$ spectrum. The noise spectrum also
shows a square root dependence on the applied power $S\sim P^{-1/2}$.
Furthermore, the recent work by Burnett et al. \cite{Burnett2014}
shows that the dependence on the applied power is also temperature
dependent. The most striking feature of the frequency noise is its
spectrum temperature dependence: $S\sim T^{-\beta}$ with ${\beta=1.2-1.73}$
\cite{Kumar2008,Burnett2013,Burnett2014} which is at odds with the
expectation that any kind of noise should disappear as $T\rightarrow0$.

On the theoretical side, the observations cannot be explained by the
conventional phenomenological model of TLS known as Standard Tunneling
Model (STM) \cite{Anderson1972,Phillips87}. This model was very successful
in explaining the anomalous bulk properties of amorphous glasses at
low temperature. However, its predictions for the frequency noise
are in a strong disagreement with the data. This problem was noted by the works \cite{Gao2008TLS,Kumar2008} which observed that
the data can be fitted by a single empirical equation that describes the noise dependence on microwave power and temperature. 
This equation however cannot be derived in the framework of STM.

In this paper we propose a model that is capable to explain all the features
of the noise. The model develops on our previous ideas \cite{Faoro2012},
it differs from other models in that it assumes relatively large interaction
between TLS. The unusual properties of the frequency noise spectra are mostly associated
with this large interaction. For a better fit to the data, the model
also assumes a slightly non-uniform density states of TLS at low energies
which might be a consequence of the large interaction. In the bulk
of the paper we show that the model explains all features of the noise
spectral density, namely, the frequency dependence of the spectrum
$S\sim f^{-1}$, the temperature dependence $S\sim T^{-\beta}$ and the applied
power dependence $S\sim P^{-1/2}$ as well as the saturation of
the noise with the power at the temperature dependent level. In addition,
the model and STM gives the same shifts in the resonant frequency as a function of temperature that were originally
interpreted as the indication for the presence of TLS \cite{Gao2008,Pappas2011}.
Thus, the predictions of the model agree well with the results of
most experiments on resonators \cite{Lindstrom2009,Wang2009,Macha2010a,Wisbey2010,Khalil2011,Sage2011}.
Furthermore, the model provides the explanation for the recent spectroscopy
data on the temperature dependence of the dephasing rate of TLS located
in the Josephson junction barriers of the phase qubits \cite{Lisenfeld2010}. 

The paper is organized as follows. Section \ref{sec:Model} gives
standard (Section \ref{sub:Standard-tunneling-model}) and generalized
(Section \ref{sub:Generalized-tunneling-model}) tunneling models
and discusses the effects of TLS interactions (Section \ref{sub:Main-predictions-of}).
The detailed calculations of the frequency noise spectrum are given
in Section \ref{sec:The-effect-of-slow-fluct} while Section \ref{sec:Discussion}
compares the predictions of the model for the noise power spectrum
with that of STM and with the experimental data. Finally, Section
\ref{sec:Conclusions} gives conclusions and discusses the possible origin
of the larger interaction assumed by the model.

\section{Model\label{sec:Model}}

\subsection{Standard tunneling model and its predictions\label{sub:Standard-tunneling-model}}

The existence of TLS in amorphous materials was conjectured four
decades ago \cite{Anderson1972,Phillips87} in order to explain the
anomalous bulk properties of these materials at low temperatures, i.e.
the temperature dependence of the specific heat and the thermal conductivity.
The phenomenological model describing the TLS is known as STM for its
simplicity and wide application. It assumes the existence of localized
excitations with very low energy $E$ that are visualized as excitations
in double well potentials that happen to be nearly symmetric. It is
generally believed that the existence of the double well potentials
is due to the disorder, so that local rearrangement of atoms might switch
the system between adjacent local energy minima. For a given $T$,
double well potentials with $E\sim k_{B}T$ dominate the thermodynamic
properties. In the double well potentials a transition between the two minima is due to the quantum tunneling. 
Therefore, they are referred as tunneling systems which are characterized by an asymmetry $\Delta$ and a tunneling
matrix element $\Delta_0$. The unperturbed Hamiltonian $H_{\text{TLS}}$
of each independent tunneling system is 
\begin{equation}
H_{\text{TLS}}=\frac{\Delta}{2}\sigma^{z}+\frac{\Delta_{0}}{2}\sigma^{x}\;
\end{equation}
Here $\sigma^{a},\, a=x,y,z$ are Pauli matrices. In the rotated basis,
the Hamiltonian is simply ${H=ES^{z}}$, where $E=\sqrt{\Delta^{2}+\Delta_{0}^{2}}$
is the TLS energy splitting and ${S^{z}=\frac{1}{2}(\cos\theta\sigma^{z}+\sin\theta\sigma^{x})}$
with $\tan\theta=\Delta_{0}/\Delta$. The STM assumes that the energy
distribution of $\Delta$ is flat while $\Delta_{0}$ is exponential
in the barrier width and thus has an exponentially wide distribution,
so that the probability density of TLS is given by 
\begin{equation}
\begin{split}P(\Delta,\Delta_{0}) & =\frac{\bar{P}_{0}}{\Delta_{0}}\Theta(\Delta_{0}-\Delta_{\text{0,min}})\\
\frac{\Delta_{\text{0,min}}}{k_{B}} & \simeq10^{-7}K
\end{split}
\label{eq:STM}
\end{equation}
The form of $P(\Delta,\Delta_{0})$ implies that the distribution
of the energy splitting $P(E)$ is uniform. Experimentally it turns
out that for most glasses $\bar{P}_{0}$ is in the range ${(0.5-3)\times10^{20}\text{eV}^{-1}\text{cm}^{-3}}$
\cite{Berret1988}.

In the insulating materials TLS are coupled to the environment by the 
interaction with phonons and photons that can excite or relax the
TLS eigenstates. The phonon interaction Hamiltonian reads:
\begin{equation}
H_{\text{TLS-ph}}=\gamma\sigma_{z}\epsilon\;\label{intph}
\end{equation}
where $\epsilon$ is the strain field and $\gamma\sim1eV$ is the
typical coupling constant. Because of this coupling the TLS acquires
a relaxation rate $\Gamma_{1}^{\text{ph}}$ and a dephasing rate $\Gamma_{2}^{\text{ph}}$.
Golden rule formula gives the relaxation rate: 
\begin{equation}
\Gamma_{1}^{\text{ph}}=\frac{\gamma^{2}}{2\pi\zeta\hbar^{4}v^{5}}\Delta_{0}^{2}E\coth[E/2k_{B}T]\label{Gamma_1^ph}
\end{equation}
where $\zeta$ the density of the glass and $v$ is the sound velocity.
The dephasing rate is due to decay: $\Gamma_{2}^{\text{ph}}=\frac{1}{2}\Gamma_{1}^{\text{ph}}$.
At low temperature, assuming that $\Delta_{0}/E$ has little or no
$E$ dependence, one concludes that $\Gamma_1\sim E^{3}$.

Because in this work we are considering TLS that are located in a very
thin layer of material on the surface of metals, we also briefly review
STM predictions for the relaxation of TLS in metals \cite{Black1977}.
In these materials, TLS also interact with the conduction electrons.
The interacting Hamiltonian reads:
\[
H_{\text{TLS-el}}=\sigma_{z}\sum_{kk'\eta}V_{kk'}c_{k\eta}^{\dagger}c_{k'\eta}
\]
where $V_{kk'}$ describes the scattering potential and ${c_{k\eta}^{\dagger}(c_{k\eta})}$
creates (annihilates) a fermion of wave vector $k$ and spin $\eta$. The coupling
between TLS and electrons is described quite general by a parameter
${\cal K}$, which for a weak s-wave potential ($V_{kk'}=V$), is ${\displaystyle {\cal K}=\frac{1}{2}(\nu_{F}V)^{2}}$,
where $\nu_{F}$ is the electron density of states at the Fermi level.
It is known that in metallic glasses ${\cal K}$ must be less than $1/2$.
This typically strong interaction leads to a short relaxation time
for the TLS: 
\begin{equation}
\Gamma_{1}^{\text{el}}=\pi {\cal K}E\coth[E/2k_{B}T]
\end{equation}
At low temperatures, $\Gamma_{1}^{\text{el}}\sim E$.

TLS can also interact between themselves due to the exchange of virtual
phonons, photons or electronic excitations. In all cases the interaction
falls off as $1/r^{3}:$ 
\begin{equation}
H_{2}^{\text{int}}=\frac{1}{2}\sum_{i,j}U_{ij}\sigma_{i}^{z}\sigma_{j}^{z}\;\quad U_{ij}=\frac{u_{ij}}{r_{ij}^{3}}\;
\end{equation}
In the case of photon exchange, the interaction is essentially the instantaneous
dipole-dipole one: 
\begin{equation}
u_{ij}=\sum_{i}\sum_{i\neq j}\frac{\vec{d}_{i}\cdot\vec{d}_{j}-3(\hat{r}_{ij}\cdot\vec{d}_{i})(\hat{r}_{ij}\cdot\vec{d}_{j})}{4\pi\varepsilon}\;
\end{equation}
In the case of phonon exchange, the retardation can be ignored if the
distance between TLS is less than the wavelength of the phonon, i.e. $r_{ij}<(vE)^{-1}$.
This condition is satisfied for characteristic distances and energies
of relevant TLS. By neglecting the retardation, the interaction is:
\begin{equation}
u_{ij}=\sum_{i}\sum_{i\neq j}\frac{\gamma_{i}\gamma_{j}}{\zeta v^{2}}\;
\end{equation}
The interaction scale is set by $U_{0}\approx d^{2}/\varepsilon$
($U_{0}\approx\gamma^{2}/\zeta v^{2}$) for electric (elastic) interactions.
Comparing the interaction between TLS at a typical distance $r^{3}\sim1/\bar{P}_{0}$
with the distance between the levels, one concludes that the effects
of the interaction are controlled by the dimensionless parameter $\chi=\bar{P}_{0}U_{0}$.
The crucial assumption of the STM is that this parameter is very small,
$\chi\ll1$, so that the effect of the interaction on TLS can be mostly
ignored. In particular, one expects that the TLS density of states remains
constant at low energies, $\rho(E)=\bar{P}_{0}$. Ultrasound attenuation
experiments that measure the product $\bar{P}_{0}U_{0}$ show that
$\chi$ is indeed small in bulk amorphous insulators and has almost
universal value ${\chi\approx10^{-3}-10^{-2}}$. In metals, the interaction
between TLS is similar to RKKY interaction between spins, so that
${\displaystyle U_{0}=E_{F}/k_{F}^{3}}$, where $E_{F}$ is the Fermi
energy and $k_{F}$ is the Fermi wave vector. In metallic glasses  $U_{0}\sim10^{5}\, K \text{\AA}^{3}$, as a result
the constant ${\chi}$ has the same order of magnitude as the phonon mediated interaction. 
To summarize, in the framework of STM the interactions of different origins
add together to form an effective interaction $U_{0}/r_{ij}^{3}$
that is characterized by the constant $U_{0}\sim10^{5}\, K \text{\AA}^{3}.$
This conclusion relies on the assumption that the TLS sizes are much smaller
than the distance between them, that allows one to estimate $r^{3}\sim1/\bar{P}_{0}$. 

The small value of the dimensionless parameter ${\chi\ll1}$ implies that the relaxation
of TLS induced by their mutual interaction is negligible. Indeed,
two interacting TLS ($i$ and $j$) exchange energy if the resonant
condition $|E_{i}-E_{j}|<U_{0}$ is satisfied. By computing the
number $N_{0}$ of TLS that form a resonant pair with a given one,
we get $N_{0}\approx\chi\ln\left(\frac{L}{a}\right)$, where $L$
is the size of the system and $a$ is the minimum distance between
two TLS \cite{Burin1996}. Because the number of resonant neighbors
$N_{0}\ll1$ for any reasonable sample size $L$, the STM assumes that
different TLS are independent and their relaxation rate $\Gamma_{1}$
is dominated by phonons.

\subsection{Generalized tunneling model\label{sub:Generalized-tunneling-model}}

In this work we show that in order to explain the data we need to
do two modifications to the standard tunneling model. We shall refer to this model as generalized tunneling model (GTM) and the two modifications
are the following:  
\begin{itemize}
\item The interaction between TLS is not neglected. In fact, we show that
the latter has a significant effect on the TLS relaxation at sufficiently low temperatures even if $\chi\ll1$. 
\item We allow a non-flat probability density of the asymmetry energy $\Delta$:
\begin{equation}
P(\Delta_{0},\Delta)=p(\Delta_{0})\left\{ \begin{array}{ll}
(1+\mu)(\frac{\Delta}{\Delta_{\text{max}}})^{\mu} & \mbox{if \ensuremath{0\leq\Delta\leq\Delta_{\text{max}}}};\\
0 & \mbox{otherwise}.
\end{array}\right.\label{eq:P(Delta_0,Delta)}
\end{equation}
where 
\begin{equation}
p(\Delta_{0})=\left\{ \begin{array}{ll}
\Delta_{0}^{-1} & \mbox{if \ensuremath{\Delta_{\text{min}}\leq\Delta_{0}\leq\Delta_{\text{max}}}};\\
0 & \mbox{otherwise}.
\end{array}\right.
\end{equation}

\end{itemize}
Here $\mu<1$ is a small positive parameter whose value will be discussed
in details below.

We notice that the second assumption might be in fact the consequence
of the first. Indeed, a strong interaction between discrete degrees
of freedom always decreases the density of states at low energies,
$\rho(E)=\rho_{0}(E/E_{\text{max}})^{\mu}$. For Coulomb interaction
this effect results in a very large suppression of the density of
states and the formation of Efros-Shklovkii pseudogap \cite{Efros1975}.
Dipole-dipole interaction is marginal and it would result in logarithmic
corrections of the density of states for point-like TLS. Because
larger than expected interaction implies that the assumption of point-like
defects is probably wrong, we do not attempt to derive the probability
distribution (\ref{eq:P(Delta_0,Delta)}) in some microscopic picture
but take it as an assumption. 

It is worthwhile to mention here that the suppression of the density of states at low energies was reported
previously by a number of experimental works. Historically, first the
specific heat measurements performed in the 80's indicated that at low temperatures
( $T\leq1\,\mbox{K}$) the density of states is $\rho(E)\sim E^{\mu}$
with $\mu\approx0.2-0.3$ \cite{Hunklinger1986}. Another indirect
evidence comes from the old flourescence experiments \cite{Selzer1976}
that showed homogenous line broadening with anomalously large magnitude
and unusual temperature dependence $\sim T^{1.3}$ in glasses. It
was argued \cite{Silbey1987} that this low temperature anomaly is
due to TLS. However, to fit the data one needs to assume a non constant
density of states $\rho(E)=\rho_{0}(E/E_{\text{max}})^{\mu}$, with
$\mu\approx0.3$. More recently, experiments by S. Skacel et al. \cite{Skacel2013} directly
probed the TLS density of states in thin a-SiO films by measuring
losses in superconducting lumped element resonators and reported that
$\rho(E)\propto E^{0.28}$ in agreement with previous measurements
in glasses.

The importance of the interaction between TLS was conjectured by Yu
and Leggett in 1988 \cite{Yu1988} who argued that the apparent universality
of the dimensionless parameter $\chi$ can be only understood as a
consequence of the many body interactions. In this picture each TLS
is a complicated many body excitation formed by many local degrees
of freedom. However, despite the effort of many workers \cite{Coppersmith1991,Burin1995b,Lubchenko2001,Vural2011,Leggett2013}
the consistent first principle theory of TLS is not available. Experimentally,
the first evidence for interactions between TLS were found in thin
$\text{a-SiO}_{2+x}$ layers, where it was shown that dipole-dipole
interactions between TLS play a key role up to 100 mk \cite{Ladieu2003}.
Very recently experiments performed on superconducting microresonators
showed that the electromagnetic response of thin oxide layers is not
described by STM. In particular, the observed weak (logarithmic) power
dependence of the loss is in a striking contrast with the square root
prediction of STM but agrees perfectly well with the interacting picture
\cite{Faoro2012}.

\subsection{Main predictions of the generalized tunneling model \label{sub:Main-predictions-of}}

Non-negligible interactions between TLS provides a mechanism for their
dephasing and relaxation that might dominate at low temperatures when
relaxation caused by phonons become very inefficient. In this section
we compute the broadening of TLS levels that is due to their mutual
interaction. Then we explain why this width is crucial for the low frequency
noise of the high quality resonators.

It is convenient to divide the TLS into coherent (quantum) and fluctuators
(classical) TLS. Coherent TLS are characterized by small phonon
induced decoherence rate, $\Gamma_{2}^{\text{ph}}<E$, while fluctuators
have $\Gamma_{2}^{\text{ph}}\geq E$. Among coherent TLS we distinguish high,
$E\gg k_{B}T$, and low, $E\lesssim k_{B}T$, energy TLS. The noise in
high quality resonators is generated by the TLS that have energies
close to the resonator frequency $\nu_{0}$. We shall assume that
the frequency of the resonator is high, $\nu_{0}\gg k_B T$, so that the TLS
responsible for the noise are high energy coherent TLS. Their properties
are affected by the environment that consists of slow fluctuators
and thermally activated coherent TLS with energies $E\lesssim k_{B}T$. 

The line width of an individual high frequency TLS is due to the combined
effect of the surrounding thermally excited TLS that change their
state emitting and absorbing phonons. We begin by evaluating the
effect of a single thermally excited TLS and then sum over many of them. 

In the rotated basis the Hamiltonian of the high frequency TLS (denoted
by subscript $0$) interacting with a thermally excited one (denoted
by subscript $T$) at distance $r$ is  given by: ${H=E_{0}S_{0}^{z}+ES_{T}^{z}+H_{\text{ph}}+H_{\text{int}}}$
with 
\begin{equation}
H_{\text{int}}=4U(r)S_{0}^{z}\left(\frac{\Delta}{E}S_{T}^{z}+\frac{\Delta_{0}}{E}S_{T}^{x}\right)\label{eq:H_int}
\end{equation}
where $U(r)=U_{0}r^{-3}$ is the interaction energy. We denote the two
states of the high frequency TLS as $\left|0\right\rangle $ and $\left|1\right\rangle $
(${S_0^{z}\left|0\right\rangle =-1/2\left|0\right\rangle}$). In the Hamiltonian (\ref{eq:H_int})
we neglected the terms proportional to $S_{0}^{x}$ that lead to decay
of the excited state. These terms are irrelevant for TLS with very
different energies, $E_{0}\gg E$. 

\begin{figure}[h]
\includegraphics[width=2.9in]{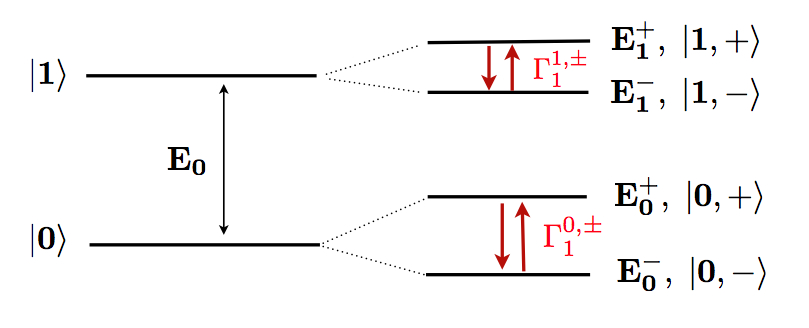} \caption{Schematics of the energy levels of the Hamiltonian $H_{\text{int}}$. The
solid arrows indicate phonon-induced relaxation.\label{fig:Schematics-of-the-energy-levels}}
\end{figure}

Hamiltonians of the type (\ref{eq:H_int}) have been studied extensively
in the context of the anomalous homogeneous optical linewidths in
glasses (\cite{Silbey1987} and refs. therein). We now outline the
main assumptions and results of these studies. Due to the interaction
the high frequency and the thermally activated TLS form a 4-levels
quantum system (see Fig. \ref{fig:Schematics-of-the-energy-levels})
which can be diagonalized by rotating the basis of thermally activated
TLS: 
\begin{equation}
\begin{split}|n,-\rangle & =\frac{1}{\sqrt{2}}\left[\sqrt{1+\eta_{n}}|n,0\rangle-\sqrt{1-\eta_{n}}|n,1\rangle\right]\\
|n,+\rangle & =\frac{1}{\sqrt{2}}\left[\sqrt{1-\eta_{n}}|n,0\rangle+\sqrt{1+\eta_{n}}|n,1\rangle\right]
\end{split}
\label{eq:rotated_basis}
\end{equation}

\begin{equation}
H_{\text{int}}=\sum_{n=0,1}\sum_{k=-,+}E_{n}^{k}|n,k\rangle\langle n,k|\;
\end{equation}
with eigenvalues: 
\begin{equation}
\begin{split}E_{0}^{\mp} & =-\frac{E_{0}}{2}\mp\sqrt{\left(\frac{E}{2}\right)^{2}+U(r)\Delta+U(r)^{2}}\\
E_{1}^{\mp} & =+\frac{E_{0}}{2}\mp\sqrt{\left(\frac{E}{2}\right)^{2}-U(r)\Delta+U(r)^{2}}
\end{split}
\label{eq:E_4_level}
\end{equation}
where
\[
{\displaystyle \eta_{n}=\frac{E+(-1)^{n}2U(r)(\Delta/E)}{\sqrt{E^{2}+(-1)^{n}4U(r)\Delta+4U(r)^{2}}}}
\]
.

The width of the sublevels of the four level system can be found by
evaluating the matrix elements describing the phonon emission or absorption
between the states (\ref{eq:rotated_basis}). A typical thermally
excited TLS is characterized by $\Delta_{0}\ll\Delta\simeq E$, so
these matrix elements are very close to the ones of non-interacting
TLS with the same energy
\[
\Gamma_{1}^{0,+}+\Gamma_{1}^{0,-}\simeq \Gamma_{1}^{1,+}+\Gamma_{1}^{1,-}\simeq\Gamma_{1}^{\text{ph}}(E).
\]

It is convenient to define the effective decoherence rate as the sum
of the widths of the sublevels weighted with their probabilities:
\begin{equation}
\Gamma_{\text{eff}}=\frac{1}{2}\sum_{k=\pm}p_{k}\left(\Gamma_{1}^{0,k}+\Gamma_{1}^{1,k}\right)\simeq\Gamma_{1}^{\text{ph}}(E)\label{eq:Gamma^eff}
\end{equation}
In the limit of significant interaction energy, $U(r)>\Gamma_{\text{eff}}$, the
width of the high frequency TLS level, $\Gamma_{2},$ coincides with the effective
rate (\ref{eq:Gamma^eff}). This can be also seen by arguing that
transition between sublevels changes the energy of the fast TLS by
$U(r)$. After such transition its wave function acquires the phase $\delta\phi=U(r)t$
and thus leads to dephasing after a time $1/U(r)$. For large $U(r)$ this
time is short compared to the time between transitions, so the dephasing
rate is given by $\Gamma_{\text{eff}}$. Note that small values of $\Gamma_{\text{eff}}\ll T$
and the fast dependence of $U(r)\sim1/r^{3}$ imply that a typical thermally
activated TLS with $U(r)>\Gamma_{\text{eff}}$ has $U(r)\ll T$. 

In the opposite limit of very small $U(r)<\Gamma_{\text{eff}}$ the phonon process
does not affect the high frequency TLS immediately. After the thermally excited
TLS changes its state, the energy of the fast TLS changes by $U(r)$,
so the phase $U(r)t$ that it acquires is much smaller than unity at
a time when the TLS flips again. As a result the effect of phonon processes
averages out. 

Both limits can be treated analytically for thermally excited TLS
with $\Delta_{0}\ll\Delta$, in which one can neglect the rotation
of the basis (\ref{eq:rotated_basis}) induced by phonon processes.
In this case the fluctuations of the TLS energy are given by 
\[
\left\langle \delta E(t)\delta E(0)\right\rangle =U(r)^{2}\cosh^{-2}(E/2T)\exp(-\Gamma_{1}^{\text{ph}}t)
\]

They result in the dephasing of the high frequency TLS 
\[
\left\langle S_{0}^{+}(t)S_{0}^{-}(0)\right\rangle \sim\left\langle \exp\left[-i\int_{0}^{t}dt_{1}\delta E(t_{1})\right]\right\rangle 
\]
In the limit ${\Gamma_{1}^{\text{ph}}t\gg1}$ the energy $\delta E(t)$ experiences many fluctuations and the average
can be evaluated in the Gaussian approximation 
\[
\left\langle S_{0}^{+}(t)S_{0}^{-}(0)\right\rangle \sim\exp \left (-\frac{u^{2}}{\Gamma_{1}^{\text{ph}}}t \right )
\]
where ${u=U(r) \cosh^{-1}(E/2T)}$. In this approximation the level width is ${\Gamma_{2}=u^{2}/\Gamma_{1}^{\text{ph}}}$.
The assumption ${\Gamma_{1}^{\text{ph}}t\gg1}$ is valid provided that ${\Gamma_{2}\ll\Gamma_{1}^{\text{ph}}}$ which is correct for ${u\ll\Gamma_{1}^{\text{ph}}}$. 

To summarize, the level width of the high frequency TLS is given by: 
\begin{equation}
\Gamma_{2}(u)=\left\{ \begin{array}{cc}
\Gamma_{1}^{\text{ph}}\, & \mbox{if} ~~u\gg\Gamma_{1}^{\text{ph}}\\
\frac{u^{2}}{\Gamma_{1}^{\text{ph}}}\, & \mbox{if} ~~ u\ll\Gamma_{1}^{\text{ph}}
\end{array}\right.\label{eq:Gamma_2(u)}
\end{equation}

The full level width of the fast TLS is given by the sum over thermally
activated TLS in its environment: 
\[
\Gamma_{2}=\sum_{k}\Gamma_{2}(u_{k})
\]
which should be averaged over positions (that control $u(r)$), energies
and relaxations rates of the thermally excited TLS. These averages
can be performed independently. Because $u\sim1/r^{3}$ the average
of (\ref{eq:Gamma_2(u)}) over positions is dominated by ${u(r)\sim\Gamma_{1}^{\text{ph}}}$.
Estimating the integral over $r$ we get 
\begin{equation}
\Gamma_{2}=c\int d\Gamma_{1}dEP(E,\Gamma_{1})U_{0}\cosh^{-1}(E/2T)\label{eq:Gamma_2_full}
\end{equation}
where $c\sim1$ and $P(E,\Gamma_{1})$ is the probability density
of TLS characterized by energy $E$ and relaxation rate $\Gamma_{1}.$
Combining the probability distribution (\ref{eq:P(Delta_0,Delta)})
and expression for the relaxation rate (\ref{Gamma_1^ph}) we get
\begin{equation}
P(E,\Gamma_{1})=P_{0}\frac{E^{\mu}}{2E_{\text{max}}^{\mu}\Gamma_{1}}\label{eq:P(E,Gamma_1)}
\end{equation}
for $\Gamma_{1}<\Gamma_{1}^{\text{max}}$ where $\Gamma_{1}^{\text{max}}=\Gamma_{1}(\Delta_{0}\sim E)$
is the maximum rate possible for TLS with energy $E$. Performing
the average in (\ref{eq:Gamma_2_full}) with the distribution (\ref{eq:P(E,Gamma_1)})
we get 
\begin{equation}
\Gamma_{2}=c\chi\ln\left(\frac{\Gamma_{1}^{\text{max}}}{\Gamma_{1}^{\text{min}}}\right)\frac{T{}^{1+\mu}}{E_{\text{max}}^{\mu}}\label{eq:Gamma_2}
\end{equation}
where $c\sim1$ and $\Gamma_{1}^{\text{min}}$ is the minimal relaxation
rate, $\ln(\Gamma^{\text{max}}/\Gamma^{\text{min}})=2\ln(E/\Delta_{0}^{\text{min}}).$
The largest value of $\Gamma^{\text{max}}$ associated with the thermally
excited TLS is of the order of $10^{7}-10^{8}\,\mbox{s}^{-1}$ for
$E \sim 11-12\,\mbox{GHz}$ \cite{Martinis2010} and correspondingly $10^{4}-10^{5}\,\mbox{s}^{-1}$ for
$T\sim50\,\mbox{mK}$. There is no information available on the precise
value of the minimal rate $\Gamma^{\text{min}}$ for thermally activated TLS in
glasses, but the electrical noise data show that $1/f$ noise generated
by these TLS extends to very low frequencies $f\lesssim1\,\mbox{mHz}$
beyond which the dependence changes. This implies that $\Gamma^{\text{min}}\lesssim10^{-3}\,\mbox{s}^{-1},$
so the value of ${\ln(\Gamma^{\text{max}}/\Gamma^{\text{min}})\approx 20}$. 

Large $\ln(\Gamma^{\text{max}}/\Gamma^{\text{min}})$ factor appears only for TLS
that are distributed uniformly through a three dimensional volume
so that the integral over the volume produces factor $U_{0}$ for any
$\Gamma_{1}$ in (\ref{eq:Gamma_2_full}). This factor is expected
to be much smaller for surface insulators. In the case of amorphous
two dimensional layers of thickness $d$ with three dimensional interaction ($U(r)\sim1/r^{3}$
) between the TLS the logarithmic contribution comes from ${\Gamma_{1}>U_{0}/d^{3}}$, which provides the lower cutoff of the
logarithmic divergence ${\Gamma_{1}^{\text{min}}\rightarrow U_{0}/d^{3}}$.
In real materials, however, the interaction between TLS might have
a two dimensional character at intermediate scales, ${d<r<d_{\text{eff}}}$
which cuts off the logarithmic divergence at smaller ${\Gamma_{1}^{\text{min}}\rightarrow U_{0}/d_{\text{eff}}^{3}}$.
For the estimates below we shall assume that $\ln(\Gamma^{\text{max}}/\Gamma^{\text{min}})\gtrsim1$
in surface oxides formed in superconducting microresonators. 

In a typical low temperature experiment the dephasing rate $\Gamma_{2}$
given by (\ref{eq:Gamma_2}) dominates over decoherence rate $\Gamma_{2}^{\text{ph}}\sim\Gamma_{1}^{\text{ph}}$
due to phonons. In fact, for $E\sim T$ we estimate the phonon mediated
relaxation rate given in (\ref{Gamma_1^ph}) 
\begin{equation}
\Gamma_{1}^{\text{ph}}\approx\frac{U_{0}}{a^{3}}\left(\frac{E}{\omega_{D}}\right)^{3}\;
\end{equation}
where $a\sim0.3\,\mbox{nm}$ is atomic distance and ${\omega_{D}=(c_{s}/a)(6\pi^{2})^{1/3}\sim10^{3}\,\mbox{K}}$
is the Debye frequency. Estimating the interaction one gets $U/a^{3}\approx300\,\mbox{K}$ \cite{Neu1997}.
A typical high frequency TLS probed by superconducting resonators or phase qubit experiments
has energy $E\sim5-10\,\text{GHz}$, for these TLS the relaxation
rate due to phonon is $\Gamma_{1}^{\text{ph}}\sim\Gamma_{2}^{\text{ph}}\sim10^{2}-10^{3}\,\text{s}^{-1}$.
At $T\sim100\,\mbox{mK}$, the dephasing rate given by (\ref{eq:Gamma_2})
is much larger: $\Gamma_{2}\sim10^{6}\,\mbox{s\ensuremath{^{-1}}}$,
assuming that $\mu\approx0.3$, $E_{\text{max}}\approx100\,\text{K}$
and $\chi\approx10^{-3}.$ Note, that the STM assumption of $\mu\approx0$
would make this rate even larger by a factor of $\sim10$. 

In contrast to the dephasing rate, the relaxation of high frequency
TLS due to interaction with others is small. The relaxation rate is
proportional to the square of the interaction, which falls of as $1/r^{6}.$
It is thus dominated by the closest TLS which is in resonance with
the given one. Because the level width of the TLS is given by $\Gamma_{2},$
the resonant condition implies that the typical distance between resonant
TLSs is $r^{3}\sim1/(\Gamma_{2}\rho(E))$, and the interaction between
them $U_{0}\Gamma_{2}\rho(E)$. Applying the Fermi-Golden rule we
estimate that the relaxation rate due to this interaction is:
\begin{equation}
\begin{split}\Gamma_{1}^{\text{TLS}} & \approx(U_{0}\rho(E))^{2}\Gamma_{2}=\chi^{2}\left(\frac{E}{E_{\text{max}}}\right)^{2\mu}\Gamma_{2}\end{split}
\label{eq:Gamma_1^TLS}
\end{equation}
The relaxation rate (\ref{eq:Gamma_1^TLS}) is much smaller than $\Gamma_{2}$
because it contains two extra factors of $\chi$ which, in contrast
to $\Gamma_{2},$ are not compensated by large logs. Estimating it
we get $\Gamma_{1}^{\text{TLS}}\sim10^{-2}-10^{0}\,\mbox{s}^{-1},$ which
is much smaller than the phonon relaxation rate. We conclude that
the phonon relaxation mechanism dominates, i.e. $\Gamma_{1}\approx\Gamma_{1}^{\text{ph}}$.

This dephasing rate (\ref{eq:Gamma_2}) is in a perfect agreement
with the direct experimental observations \cite{Lisenfeld2010}
that used phase qubits to study individual TLS with energies ${E\sim6-8\,\mbox{GHz}}$.
This work observed the temperature dependence ${\Gamma_{2}\propto T^{1.24}}$
and absolute values ${\Gamma_{2}\sim10^{6}\,\mbox{s}^{-1}}$ at ${T\sim50\,\mbox{mK}}$.

The discussion above does not differentiate between coherent and incoherent
thermally excited TLS. The small fluctuations of the energy of the
high frequency TLS created by coherent and incoherent TLS far away
from the fast TLS are indistinguishable. The crucial assumption in
the derivation of the level width $\Gamma_{2}$ (\ref{eq:Gamma_2})
was the guassian nature of the effective energy fluctuations $\delta E$
which is the sum of the effects produced by many flucutuators. This
assumption is confirmed by the large factor $\ln(\Gamma_{1}^{\text{max}}/\Gamma_{1}^{\text{min}})$
that appeared in (\ref{eq:Gamma_2}). 

\begin{figure}[h]
\includegraphics[width=2.5in]{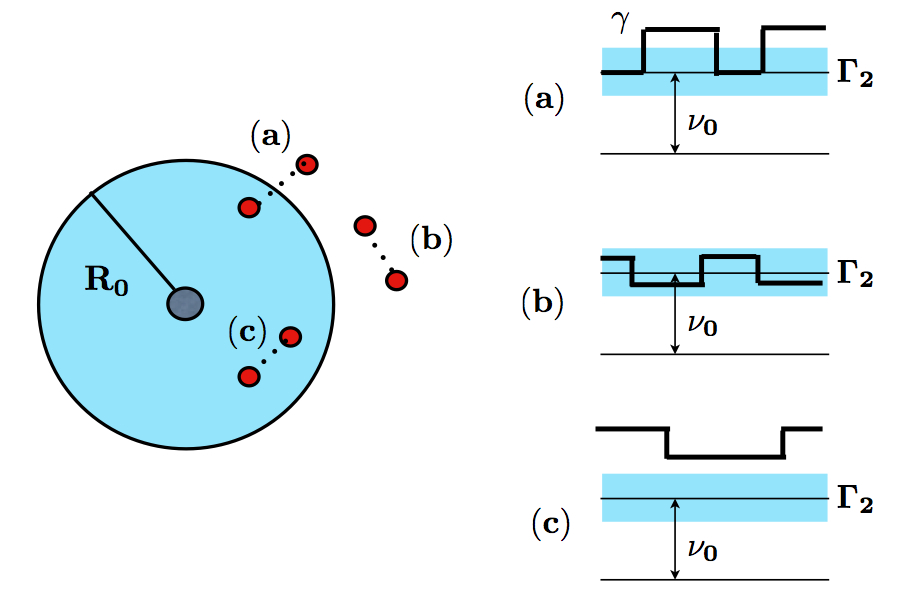} \caption{High frequency TLS (dark small circle) and  fluctuators that are coupled
to it. The strongly coupled fluctuator (a) brings the TLS in and out of
resonance with the external proble. This translates into a large noise
measured by the probe. The weakly coupled fluctuator (b) only contributes
to the line width of the high frequency TLS. The strongly coupled fluctuator (c) is not strong enough
to bring the fast TLS in the resonance, so its effect is not observable.
\label{fig:High-Frequency-TLS}}
\end{figure}

The effect of the slow fluctuators requires a separate analysis for
those fluctuators that are located so close to the high frequency TLS that
they shift its energy by an amount larger than the width $\Gamma_{2}$.
As mentioned above, the presence of slow fluctuators is revealed
by the omnipresent $1/f$ charge and critical current noise that extends
to the lowest frequencies \cite{Weissman1988}. Some of these fluctuators
interact strongly with the fast TLS: $U(r)>\Gamma_{2}$ for $r<R_{0}$,  
where ${R_0^3 = U_0/\Gamma_2}$.
These fluctuators create highly non-gaussian noise that cannot be
regarded as a contribution to $\Gamma_{2}.$ Qualitatively, the slow
strong fluctuators result in the chaotic motion of individual TLS
levels around their average positions as shown in Fig. \ref{fig:High-Frequency-TLS}
where we sketch the effect of different fluctuators on high frequency
TLS. Strongly coupled fluctuators (a) are located within the sphere
of radius $R_{0}$ and brings TLS in and out of resonance with the external
probe. The fluctuator (b) is weakly coupled and contributes to the level
width. The fluctuator (c), although strong enough to be non-gaussian,
is not sufficiently strong to bring the TLS in resonance with the external
probe. The chaotic motion of TLS energy level due to the strong fluctuators
causes the noise in the external probe, such as resonator frequency.
We discuss this noise in the following Section. 

\section{The effect of slow fluctuators on the resonator noise\label{sec:The-effect-of-slow-fluct}}

\begin{figure}[h]
\includegraphics[width=3.5in]{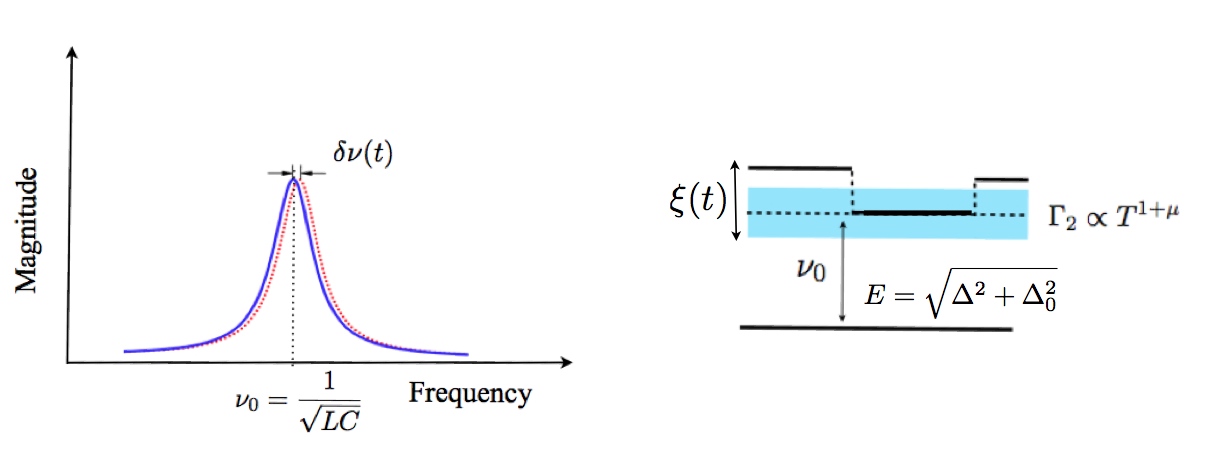} \caption{Schematics of the frequency noise generation in microresonators. The
noise is due to fluctuators that are strongly coupled to resonant
TLS and can induce energy drifts for the resonant TLS larger than
the broadening width $\Gamma_{2}$ by bringing the resonant TLS in
and out of resonance with the resonator. \label{fig:Schematics-of-the-frequency-noise}}
\end{figure}

The frequency noise in the microresonator is ultimately due to the
switching of classical fluctuators that are strongly coupled to TLS
that are in resonance with the resonator electromagnetic mode. The
coupling is strong in the sense that the resulting energy drift of
the resonant TLS is larger than the broadening of its level, $\Gamma_{2}$, i.e. 
$U(r)>\Gamma_{2}$. The condition $U(r)>\Gamma_{2}(T)$ is satisfied
for all fluctuators in the sphere of radius $R_0$ around the resonant TLS.
Because the width $\Gamma_{2}(T)$ decreases at low temperatures,
the volume of the sphere of radius $R_0$ grows at low temperatures.
This compensates the decrease in the density of thermally activated
fluctuators. The effect of each TLS on the dielectric constant and
thereby on the resonator frequency is proportional to $1/\Gamma_{2}$.
Thus, as the temperature goes down, the noise increases: a conclusion
that seems to contradict the intuition. We illustrate the mechanism
of the resonator noise in Fig. (\ref{fig:Schematics-of-the-frequency-noise}).
The motion of levels in and out of the resonance does not affect the
average dielectric constant of the material because the average number
of TLS in resonance with the external frequency remains the same.
Thus, one expects that in contrast to temperature dependent noise,
neither internal loss nor average frequency shift of the resonators
show anomalous temperature dependence. 

The classical fluctuators responsible for the effects discussed in
this section might be slow TLS that are characterized by small $\Delta_{0}$
and $\Gamma_{1}$ or have a different nature. The main results of
the following discussion do not depend on the assumption that classical
fluctuators have the same nature as TLS, but when estimating the magnitude
of the effect we shall assume that they have similar densities. 

We now provide the detailed computation that confirms this qualitative
conclusion and provides quantitative estimates of the noise. The interaction
between TLS and electrical field, $\vec{{\cal E}}(t)=\vec{{\cal E}}\cos\nu_{0}t$
in the resonator is due to its dipole moment, $\vec{d}_{0}$ :

\begin{equation}
H_{\text{field}}^{\text{int}}=\vec{d}_{0}\cdot\vec{{\cal E}}(t)\sigma^{z}.
\end{equation}

The dynamics of the coherent TLS can be described by the Bloch equations
\cite{Bloch1957}, which coincide with the equation for the TLS density
matrix evolution. These equations includes the phenomenological description
of the decay and decoherence process with rates $\Gamma_{1}$ and
$\Gamma_{2}$. The effect of the classical fluctuators is described
by an additional time dependent contribution to the effective 'magnetic'
field acting on the pseudospin representing the TLS: $\vec{B}(t)=\vec{B}'+\vec{B}''(t)$, 
where $\vec{B}'=(0,0,E-\xi(t))$ and $\vec{B}''(t)=2(\sin\theta,0,\cos\theta)\vec{d_{0}}\cdot\vec{{\cal E}}(t)$.
The ac electric field $\vec{{\cal E}}(t)=\vec{{\cal E}}\cos\nu_{0}t$
is a small perturbation, so one can linearize the Bloch equations
by keeping terms of the first order in the applied electric field.
We look for solutions of the Bloch equations of the form $\vec{S}(t)=\vec{S}^{0}(t)+\vec{S}^{1}(t)$,
where $S^{0}$ is the solution in the absence of electric field and
$S^{1}\propto\vec{{\cal E}}(t)$. The linearized equations become
\begin{equation}
\begin{split}\frac{dS_{z}^{0}(t)}{dt} & ={\cal I}mS^{+}(t)\Omega\cos\nu_{0}t-\Gamma_{1}^{\text{ph}}\left[S_{z}^{0}(t)-m\right]\\
i\frac{dS^{+}(t)}{dt} & =\left[E+\xi(t)-i\Gamma_{2}\right]S^{+}(t)+\Omega S_{z}^{0}(t)\cos\nu_{0}t
\end{split}
\label{eq:Bloch_eqs}
\end{equation}
Here we have introduced the raising operator ${S^{+}=S_{x}^{1}+iS_{y}^{1}}$,
${\Omega=2\sin\theta\vec{d}_{0}\cdot\vec{{\cal E}}}$ is the Rabi frequency
and ${m=\tanh(E/2k_{B}T)/2}$. The presence of fluctuators (weakly and
strongly coupled to the TLS) is accounted by the energy drift $\xi(t)$. 

The physical quantity that we need to get from the solution of (\ref{eq:Bloch_eqs})
is the average polarization ${{\bf P}_{\nu_{0}}(t)}$ produced by the
resonant TLS: 
\begin{equation}
{\bf P}_{\nu_{0}}(t)=\frac{1}{2}\langle\vec{d}_{0}\sin\theta\langle S^{+}(t)\rangle_{f}\rangle=\varepsilon\chi(\nu_{0},t)\vec{{\cal E}}\label{pol}
\end{equation}
where $\langle\cdot\rangle_{f}$ denotes the average over the distribution
of the strongly coupled fluctuators responsible for the energy drift and
the average $\langle\cdot\rangle$ is taken over the distribution
of all the coherent TLS and their dipole moments. The coefficient
$\chi(\nu_{0},t)$ gives the permittivity which is responsible for
the variation of the complex resonance frequency \cite{Kumar2007}:
\begin{equation}
\frac{\delta f^{*}}{f^{*}}=-\int_{V_{h}}\frac{\chi(\nu_{0},t)|\vec{{\cal E}}|^{2}dV}{2\int_{V}|\vec{{\cal E}}|^{2}dV}\label{eq:cshift}
\end{equation}
where $V_{h}$ is the TLS host material volume and $V$ is the resonator
volume. The real part of (\ref{eq:cshift}) gives the relative frequency
shift 
\begin{equation}
\frac{\delta\nu(t)}{\nu_{0}}=-\frac{\int_{V_{h}}{\cal R}e[{\bf P}_{\nu_{0}}(t)]\cdot\vec{{\cal E}}dV}{2\varepsilon\int_{V}|\vec{{\cal E}}|^{2}dV}\label{eq:shift}
\end{equation}
while the imaginary part is responsible for the internal quality factor
$Q$: 
\begin{equation}
\left(\frac{1}{Q}-\frac{1}{Q_{0}}\right)=\frac{\int_{V_{h}}{\cal I}m[{\bf P}_{\nu_{0}}(t)]\cdot\vec{{\cal E}}dV}{2\varepsilon\int_{V}|\vec{{\cal E}}|^{2}dV}\label{loss}
\end{equation}
The frequency noise spectrum measured in the microresonator is defined
as: 
\begin{equation}
\frac{S_{\delta\nu}}{\nu_{0}^{2}}=\lim_{\tau\to\infty}\frac{1}{\tau}\int_{0}^{\tau}\int_{0}^{\tau}\frac{\langle\delta\nu(t_{1})\delta\nu(t_{2})\rangle}{\nu_{0}^{2}}e^{i\omega(t_{1}-t_{2})}dt_{1}dt_{2}\label{eq:noise}
\end{equation}
Notice that both the frequency shifts and the noise are related to
the real part of the susceptibility.

Our goal is to get the physical quantities (\ref{eq:shift}-\ref{eq:noise})
from the solution of the Bloch equation (\ref{eq:Bloch_eqs}). The
assumption that relevant fluctuators are slow, implies that we can
solve the equations (\ref{eq:Bloch_eqs}) in the stationary approximation:
\begin{equation}
{\cal R}eS^{+}(t)=\frac{\Omega m\left[\nu_{0}-E-\xi(t)\right]}{\left[\nu_{0}-E-\xi(t)\right]^{2}+\Gamma_{2}^{2}+\Omega^{2}\Gamma_{2}(\Gamma_{1}^{\text{ph}})^{-1}}\label{eq:stat_sol}
\end{equation}
In order to calculate the average polarization $P_{\nu_{0}}(t)$ given
by (\ref{pol}) we need to average (\ref{eq:stat_sol}) first over
the distribution of fluctuators and then over the distribution of coherent
TLS.

Generally, the energy drift caused by fluctuators can be written as
$\xi(t)=\sum_{k}^{{\cal N}_{f}}u_{k}n_{k}(t)$, where ${\cal N}_{f}$
is the number of coupled fluctuators, $u_{k}=U_{0}/r_{k}^{-3}$
denotes the interaction strength of the $k$-th fluctuator coupled
to the resonant TLS and $n_{k}(t)=\pm1$ is a random telegraph signal
with associated switching rate $\gamma_{k}$. Effectively each fluctuator
produces a random telegraph signal with the following properties:\\
 - $n_{k}(t)=\pm1$ with probabilities $p(n_{k}(0)=\pm1)=1/2$;\\
 - the number ${N_{*}}$ of zero crossings in the interval ${(0,t)}$
is described by a Poisson process with probabilities: 
\[
\left\{ \begin{array}{ll}
p(N_{*}=\text{even number})=e^{-\gamma_{k}t}\cosh\gamma_{k}t\\
p(N_{*}=\text{odd number})=e^{-\gamma_{k}t}\sinh\gamma_{k}t\;
\end{array}\right.
\]
We now show that weakly coupled fluctuators do not contribute neither
to the frequency noise or the frequency shifts because their contribution
to the real part of the response is equivalent to a mere additional
broadening for the resonant TLS. The solution (\ref{eq:stat_sol})
implies that in this computation we can neglect the time dependence
of $\xi(t)$, which we emphasize by writing its argument as the subscript
$\xi(t)=\xi_{t}$. In order to average over weakly coupled fluctuators
the real part of the response
\begin{equation}
\langle{\cal R}eS^{+}(t)\rangle_{f}=\int{\cal R}eS^{+}(t)P_{{\cal N}_{f}}(\xi_{t})d\xi_{t}\label{eq:S^+ave}
\end{equation}
we need to compute the distribution $P_{{\cal N}_{f}}(\xi_{t})$
defined by 
\begin{equation}
P_{{\cal N}_{f}}(\xi_{t})=\prod_{k=1}^{{\cal N}_{f}}\left[\int dz_{k}P(z_{k})\right]\delta\left(\sum_{k'=1}^{{\cal N}_{f}}z_{k'}-\xi_{t}\right)\;\label{dist}
\end{equation}
where $z_{k}=u_{k}n_{k}$ and $P(z_{k})$ is the distribution of the
$k$-th RTS. The constraint imposed by the $\delta-$function can be simplified
by finding first the Fourier transform, ${G_{{\cal N}_{f}}(\lambda)=\int P_{{\cal N}_{f}}(\xi)\exp \left [i\lambda\xi\right ] d\xi}$:
\begin{equation}
\begin{split}G_{{\cal N}_{f}}(\lambda) & =\left[\int_{-\infty}^{\infty}e^{i\lambda z_{k}(t)}P(z_{k})dz_{k}\right]^{{\cal N}_{f}}\\
 & =\left[\frac{1}{V_{h}}\int dr_{k}^{3}\cos\left(\frac{U_{0}\lambda}{r_{k}^{3}}\right)\right]^{{\cal N}_{f}}
\end{split}
\label{eq:G(lambda)_def}
\end{equation}
here $V_{h}$ is the fluctuators host material volume. Integrating (\ref{eq:G(lambda)_def})
we find: 
\begin{equation}
\begin{split}\, G_{{\cal N}_{f}}(\lambda) & =\exp\left\{ \frac{{\cal N}_{f}}{V_{h}}\int dr_{k}^{3}\left[\cos\left(\frac{U_{0}\lambda}{r_{k}^{3}}\right)-1\right]\right\} \\
 & =\exp\left[-\Gamma_{f}|\lambda|\right]
\end{split}
\label{eq:G(lambda)}
\end{equation}
where ${\Gamma_{f}=C\rho_{0f}U_{0}}$, ${\rho_{0f}\approx\rho_{0}T^{1+\mu}/E_{max}^{\mu}}$
is the density of thermally activated fluctuators and ${C=\frac{4\pi}{3}\int_{0}^{\infty}dy\left(1-\cos\frac{1}{y}\right)\approx6.57}$
is a constant. By performing the inverse Fourier Transform of Eq.
(\ref{eq:G(lambda)}) we get the distribution 
\begin{equation}
\begin{split}P(\xi(t)) & =\int_{-\infty}^{+\infty}d\lambda e^{-i\lambda\xi(t)-\Gamma_{f}|\lambda|}\\
 & =\sqrt{\frac{2}{\pi}}\frac{\Gamma_{f}}{\Gamma_{f}^{2}+\xi(t)^{2}}\;
\end{split}
\label{eq:P_Nf(xi)}
\end{equation}
that is Lorentzian. By substituting (\ref{eq:stat_sol}) and (\ref{eq:P_Nf(xi)})
into (\ref{eq:S^+ave}) we get the response: 
\begin{eqnarray}
\langle{\cal R}eS^{+}(t)\rangle_{f} & = & \frac{\sqrt{2\pi}\Omega m\left(\nu_{0}-E\right)}{\left(\nu_{0}-E\right)^{2}+\left(\sqrt{\Gamma_{2}^{2}+\Omega^{2}\Gamma_{2}(\Gamma_{1}^{\text{ph}})^{-1}}+\Gamma_{f}\right)^{2}}\;
\notag
\end{eqnarray}
that shows the additional contribution, $\Gamma_{f}$, to the dephasing
width. Unlike $\Gamma_{2}$ (\ref{eq:Gamma_2}) this contribution
does not contain a large logarithmic factor, so $\Gamma_{f}\ll\Gamma_{2}$
for bulk materials. As explained in section \ref{sub:Main-predictions-of}
the logarithmic factor might become of the order of unity for surface
dielectrics, so in this case $\Gamma_{f}\lesssim\Gamma_{2}$. 

We now discuss the effect of strongly coupled fluctuators on the real part of the susceptibility.
Estimating the number of strongly coupled fluctuators by ${{\cal N}_{f}=\frac{4\pi}{3}\frac{U_{0}}{\Gamma_{2}}\rho_{0}}$
we get ${{\cal N}_{f}\sim1}$ for two dimensional surface dielectrics and ${{\cal N}_{f}\sim10^{-1}}$
for three dimensional materials characterized by a large value of
$\ln(\Gamma_{1}^{\text{max}}/\Gamma_{1}^{\text{min}})$. The same estimate can be
obtained directly from the experimental values ${\rho_{0}\approx10^{20}\text{cm}^{-3}\text{eV}^{-1}}$,
${\Gamma_{2}\approx 2\cdot10^{-4}\text{K}}$ and ${U_{0}\approx10\text{K}\text{nm}^{3}}$.
A strongly coupled fluctuator brings the resonant TLS in and out
of resonance inducing a dynamical change of the susceptibility that
is described by a random telegraph signal: 
\begin{equation}
\begin{split}w_{k}^{\text{res}} & =\lim_{u_{k}\to0}\frac{\Omega m\left[\nu_{0}-E-u_{k}\right]}{\left[\nu_{0}-E-u_{k}\right]^{2}+\Gamma_{2}^{2}+\Omega^{2}\Gamma_{2}(\Gamma_{1}^{\text{ph}})^{-1}}\\
w_{k}^{\text{off}} & =0\;
\end{split}
\label{eq:w_k}
\end{equation}
with the switching rate $\gamma_{k}$ of the strongly coupled fluctuator.
As a result, it contributes to the the average susceptibility as $w_{k}^{\text{res}}(\tanh E/2T+1)/2$.
By substituting (\ref{eq:w_k}) into (\ref{eq:shift}) we estimate
the induced frequency shift of the resonator: 
\begin{eqnarray}
\frac{\delta\nu}{\nu_{0}}=\frac{1}{3}\langle\vec{d}_{0}^{2}\rangle\frac{\int_{V_{h}}v(\nu_{0},\vec{{\cal E}},T)|\vec{{\cal E}}|^{2}dV}{2\int_{V}\epsilon|\vec{{\cal E}}|^{2}dV}\;\label{shifts0}
\end{eqnarray}
where 
\begin{equation}
v(\nu_{0},\vec{{\cal E}},T)=\int_{0}^{E_{max}}dE\frac{P(E)\tanh\left(\frac{E}{2T}\right)(E-\nu_{0})}{(E-\nu_{0})^{2}+\Gamma_{2}^{2}+\Omega^{2}\Gamma_{2}(\Gamma_{1}^{\text{ph}})^{-1}}\;\label{intp}
\end{equation}
Notice that the frequency shift given by (\ref{shifts0}) is very
similar the ones predicted by  the STM. The only difference between (\ref{shifts0})
and the STM predictions is associated with the different probability distribution
assumed for the energy splitting of the resonant TLS but the shifts
are completely insensitive to the presence of weakly and strongly
fluctuators coupled to resonant TLS. As a result, the presence of
strongly interacting fluctuators cannot be detected by the measurements
of the frequency shifts as a function of temperature. However, as we have already shown in a previous
work \cite{Faoro2012} the presence of fluctuators is revealed by
the power dependence of the losses in high quality microresonator.
The fluctuators result indeed in a weaker (logarithmic) dependence of the losses
on the applied power which is in very good agreement with data, in
contrast with the square root dependence predicted by the STM theory \cite{Pappas2011,Macha2010a,Khalil2011}.

We now demonstrate that interaction between resonant TLS and strongly coupled fluctuators
affects significantly the noise in microresonator. As it is evident
from (\ref{eq:noise}), the noise spectrum of the microresonator is
the Fourier transform of the autocorrelation function of the susceptibility.
Each fluctuator that is strongly coupled to a resonant TLS contributes
to the autocorrelation function of the susceptibility as: \ensuremath{\frac{1}{4}\left(w_{k}^{\text{res}}-w_{k}^{\text{off}}\right)^{2}e^{-2\gamma_{k}(t_{2}-t_{1})}}
 and consequently to the noise spectrum of the microresonator as a
Lorentzian. By summing over different TLS coupled to strongly coupled
fluctuators we find that that the noise spectrum is:
\begin{equation}
\frac{S_{\delta\nu}}{\nu_{0}^{2}}(\omega)=\frac{8}{15}\langle\vec{d}_{0}^{4}\rangle \mathcal{P}(\nu_{0},\vec{{\cal E}},T)\int\frac{\gamma P(\gamma)}{\gamma^{2}+\omega^{2}}d\gamma\label{noise0}
\end{equation}
Here $P(\gamma)$ is the probability distribution of the switching
rates of the strongly coupled fluctuators, 
\[
\mathcal{P}(\nu_{0},\vec{{\cal E}},T)=\frac{\int_{V_{h}}s(\nu_{0},\vec{{\cal E}},T)|\vec{{\cal E}}|^{4}dV}{4\left(\int\epsilon|\vec{{\cal E}}|^{2}dV\right)^{2}}
\]
depends on the volume $V_{h}$ taken by the amorphous material and 
\begin{equation}
s(\nu_{0},\vec{{\cal E}},T)=\int\frac{(\nu_{0}-E)^{2}\tanh^{2}\left(\frac{E}{2T}\right)dEP(E)}{\left[(\nu_{0}-E)^{2}+\tilde{\Gamma}_{2}^{2}+\Omega^{2}\tilde{\Gamma}_{2}(\Gamma_{1}^{\text{ph}})^{-1}\right]^{2}}\;\label{int}
\end{equation}
which depends on the temperature and the power applied to the microresonator.

The frequency dependence of the noise spectrum given in  (\ref{noise0})
is $1/f$ if the switching rate $\gamma$ has $P(\gamma)\sim1/\gamma$
distribution. Such distribution is expected for practically all realistic
models of fluctuators. For instance, fluctuators that represent slow
TLS flipped by phonons, has $P(\Gamma_{1})\sim1/\Gamma_{1}$ as explained
in Section \ref{sub:Main-predictions-of}. More generally, any process which rate depends exponentially
on a physical quantity, $l$, with a smooth distribution is characterized
by $P(\gamma)\sim1/\gamma$ distribution in the exponentially wide
range $\gamma_{\text{min}}\ll\gamma\ll\gamma_{\text{max}}$. For instance,
such distribution for the switching rate appears for a particle
trapped in a double-well potential whose quantum tunneling rate through
the potential barrier depends exponentially on both the height and
the width of the barrier, as well as for a thermally activated tunneling
with rate ${\gamma_{0}e^{-E_{a}/K_{B}T}}$, where ${E_{a}}$ denotes the 
activation energy.

The dependence of the noise spectrum on the temperature and the power
applied to the microresonator can be found by performing the integral
given in (\ref{int}). The result has different structure at low and
high temperature. Because $\sqrt{\tilde{\Gamma}_{2}^{2}+\Omega^{2}\tilde{\Gamma}_{2}(\Gamma_{1}^{\text{ph}})^{-1}}\ll\nu_{0}$,
at low temperature $T\ll\nu_{0}$ the integral is dominated by small vicinity of $\nu_{0}$: 
\begin{equation}
\begin{split}s(\nu_{0},\vec{{\cal E}},T) & \simeq\left(\frac{\nu_{0}}{E_{\text{max}}}\right)^{\mu}\frac{\bar{P}_{0}}{\Gamma_{2}\sqrt{1+|\vec{{\cal E}}/{\cal E}_{c}|^{2}}}\end{split}
\label{eq:s(nu,E,T)_lowT}
\end{equation}
where ${\cal E}_{c}$ has a physical meaning of the critical field for the TLS saturation. It is defined by 
\begin{equation}
{\cal E}_{c}=\frac{\sqrt{\Gamma_{1}^{\text{ph}}\Gamma_{2}}}{2\langle\vec{d}_{0}\left|\sin\theta\right|\rangle}
\end{equation}
The important property of the generalized tunneling model is that
the critical electric field ${\cal E}_{c}$ is temperature dependent
and it scales as as ${\cal E}_{c}\propto T^{\frac{1+\mu}{2}}$.

By substituting (\ref{eq:s(nu,E,T)_lowT}) into (\ref{noise0}), we
find that in the low temperature limit the noise spectrum is 
\begin{equation}
\frac{S_{\delta\nu}}{\nu_{0}^{2}}(\omega)\sim\frac{\chi}{\omega}\left(\frac{\nu_{0}}{E_{\text{max}}}\right)^{\mu}\frac{U_{0}}{\Gamma_{2}}\left\{ \begin{array}{ll}
\frac{\int_{V_h} { \cal E}_c  |\vec{{\cal E}}|^3 dV}{4 \left ( \int_V \epsilon |\vec{{\cal E}}|^2 dV \right )^2}  & \mbox{ if \ensuremath{|\vec{{\cal E}}|\gg{\cal E}_{c}}};\\
\frac{\int_{V_h} |\vec{{\cal E}}|^4 dV}{4 \left ( \int_V \epsilon |\vec{{\cal E}}|^2 dV \right )^2} & \mbox{ if \ensuremath{|\vec{{\cal E}}|\ll{\cal E}_{c}}}.
\end{array}\right.\label{eq:noise_lowT}
\end{equation}

At all radiation powers the spectrum of the noise has $1/f$ dependence.
In a strong electric field the spectrum scales with the applied power
as $\sim P^{-1/2}$ and with temperature as $\sim T^{(1-\mu)/2}$
while in the weak electric field regime it is power independent and
scales with temperature as $\sim T^{-(1+\mu)}$.

At high temperatures, $T\gg\nu_{0}$ the $1/f$ frequency dependence
of the noise power remains intact but its temperature dependence changes.
Evaluating the integral (\ref{int}) in this limit we find 
\begin{equation}
\begin{split}s(\nu_{0},\vec{{\cal E}},T) & \simeq c\bar{P}_{0}\frac{T{}^{\mu-1}}{E_{max}^{\mu}}\end{split}
\label{eq:s(nu,E,T)_highT}
\end{equation}
where ${\displaystyle c=\int_{0}^{\infty}dxx^{\mu-2}\tanh^{2}(x/2)\approx1.2}$.
By substituting (\ref{eq:s(nu,E,T)_highT}) into (\ref{noise0}) we
find the noise spectrum in this regime 
\begin{equation}
\frac{S_{\delta\nu}}{\nu_{0}^{2}}(\omega)\sim\frac{\chi}{\omega}\frac{U_{0}}{T}\left(\frac{T}{E_{\text{max}}}\right)^{\mu}\frac{\int_{V_h} |\vec{{\cal E}}|^4 dV}{4 \left ( \int_V \epsilon |\vec{{\cal E}}|^2 dV \right )^2} \label{eq:noise_highT}
\end{equation}
In this regime the noise spectrum has weaker temperature dependence,
$\sim T^{\mu-1}$ and has no power dependence.

In the intermediate temperature $T\sim\nu_0$, one expects a smooth
crossover between the limits (\ref{eq:noise_lowT}) and (\ref{eq:noise_highT}), leading to predictions for the noise spectrum that is in agreement
with the data.

\section{Discussion\label{sec:Discussion}}

The theoretical expectations derived in the previous section, are
in a very good agreement with main features of the data
\cite{Gao2007,Kumar2007,Zmuidzinas2012,Burnett2013,Murch2012,Neill2013,Burnett2014}. 
Most importantly the equations (\ref{eq:noise_lowT}) and (\ref{eq:noise_highT})
give correct power and temperature dependence of the noise spectra.
In particular, these spectra display the very unusual behavior, observed
experimentally, of the noise increasing at low temperatures. There
is no contradiction between this growth and the Nernst Theorem, because
it is due to the fact that the sensitivity of individual TLS to the
slow fluctuators increases dramatically at low temperatures. 

The growth of the noise at low temperatures is a clear evidence of
the importance of the interactions between TLS. Indeed, the STM gives
completely different predictions for the temperature dependence of
the noise, as we show now. We focus on the weakly driven TLS in which
computations are straightforward. The Bloch equations (\ref{eq:Bloch_eqs})
become 
\begin{equation}
\begin{split}\frac{dS_{z}^{0}(t)}{dt} & =-\Gamma_{1}\left[S_{z}^{0}(t)-m\right]\;\\
i\frac{dS^{+}(t)}{dt} & =\left[E-i\Gamma_{2}\right]S^{+}(t)+\Omega S_{z}^{0}(t)\cos\nu_{0}t\;
\end{split}
\end{equation}
which solutions are: 
\begin{eqnarray}
S_{z}^{0}(t) & = & m+\left[S_{z}^{0}(0)-m\right]e^{-\Gamma_{1}t}\\
S^{+}(t) & = & \frac{\Omega\left[(E-i\Gamma_{2})\cos\nu_{0}t-i\omega\sin\nu_{0}t\right]m}{\nu_{0}^{2}-(E-i\Gamma_{2})^{2}}\label{eq:S^+(t)}\\
 & + & \frac{\Omega\left[(E-i\Gamma_{+})\cos\nu_{0}t-i\nu_{0}\sin\nu_{0}t\right]\delta S_{t}^{z}}{\nu_{0}^{2}-(E-i\Gamma_{+})^{2}}\nonumber 
\end{eqnarray}
where $\delta S_{t}^{z}=\left[S_{z}^{0}(0)-m\right]e^{-\Gamma_{1}t}$
and $\Gamma_{+}=\Gamma_{1}+\Gamma_{2}$. The first term in (\ref{eq:S^+(t)})
describes the average response, the second the relaxation after the
spin flip process which is responsible for the noise. Because the
frequency shift of the resonator is due to ${\langle \cal R}eS^{+}(t)\rangle$, the
noise in this quantity is given by $\langle{\cal R}eS^{+}(t){\cal R}eS^{+}(0)\rangle$
which is proportional to $\left\langle \left(S_{z}(0)-m\right)^{2}\right\rangle =1-m^{2}=\cosh^{-2}(E/2T)$.
In the low temperature regime $T\ll\nu_{0}$, at relevant energies
$E\ll\nu_{0}$ and $\Gamma\ll E$, we find that the noise spectrum is 
\begin{equation}
\begin{split}\frac{S_{\delta\nu}}{\nu_{0}^{2}}(\omega) & \sim\frac{\bar{P}_{0}}{\omega}\int\frac{E^{2}dE}{\nu_{0}^{4}\cosh^{2}E/2T}\frac{\int_{V_h} |\vec{{\cal E}}|^4 dV}{4 \left ( \int_V \epsilon |\vec{{\cal E}}|^2 dV \right )^2} \\
 & \propto\left[\bar{P}_{0}V_{h}T\right]\frac{T^{2}}{\omega}={\cal N}_{TLS}\frac{T^{2}}{\omega}
\end{split}
\end{equation}
where ${\cal N}_{TLS}$ is the number of thermally activated TLS
located in the dielectric volume $V_{h}$. Although the noise spectrum
has the correct, $1/f$, frequency dependence, its power decreases
quickly at low temperatures in a sharp contrast to the data.

\section{Conclusions\label{sec:Conclusions}}

The predictions of the generalized tunneling model for the noise spectra
of the resonator frequency derived in the previous sections agree very well with
recent detailed measurements performed in high-Q superconducting microresonators
\cite{Burnett2014}. Reversing the logic one can extract the phenomenological
parameter $\mu$ from these data. The resulting value $\mu\approx0.2-0.4$
is in a very good agreement with the value that was found in many
bulk glasses \cite{Silbey1987}. This value agrees perfectly well with the direct measurements of
the dephasing rate of TLS in the insulating barrier of phase qubits
that give $\Gamma_{2}\propto T^{1+\mu}$ with $\mu\approx0.24$ \cite{Lisenfeld2010}.
The absolute values of the dephasing rate observed in these experiments
agree well with the theoretical expectations assuming $\chi(T/E_{max})^{\mu}\sim10^{-3}$. 

As was emphasized repeatedly by Leggett the apparent universality
of the dimensionless parameter $\chi\sim10^{-3}$ in the STM is very
strange and asks for theoretical explanation. In the GTM considered
in this paper this puzzle becomes less striking because the parameter
that controls the interaction between the TLS has a weak energy dependence:
$\chi_{\text{eff}}=\chi(T/E_{max})^{\mu}$. At low temperatures $T\sim100\,\mbox{mK}$
this parameter becomes much smaller than its high energy (bare) value.
Assuming that the power law $(E/E_{max})^{\mu}$ extends to the atomic
energy scales, $E_{max}\sim10^{3}\,\mbox{K}$, one deduces the \emph{bare}
value of the parameter $\chi_{0}\sim10^{-1}-10^{-2}$. The fact that
the value of $\chi_{0}$ at high temperature is somewhat small is
not surprising, because larger values would imply melting. Indeed,
the average thermal displacements, $\delta u$, of all TLS per atomic
volume is $\left\langle \delta u^{2}\right\rangle _{\text{Th}}\sim d^{2}TP_{0}a^{3}$
where $a$ is interatomic spacing and $d$ is a typical displacement
caused by TLS. The Lindemann melting criterion demands that $\left\langle \delta u^{2}\right\rangle _{\mbox{Th}}<(c_{L}a)^{2}$
where $c_{L}\approx0.1-0.2$ is Lindemann parameter. Estimating the
interaction parameter $U_{0}\sim\omega_{D}d^{2}a$ we can rewrite
the Lindemann melting condition as $(T/\omega_{D})\chi_{0}<c_{L}^{2}$
which implies that the maximal values of $\chi_{0}$ consistent with
the glass stability are $\chi_{0}\sim10^{-1}-10^{-2}$. 

In conclusion, the data and their theoretical analysis remove the
mystery of the universality of the dimensionless parameter ${\chi\sim10^{-3}-10^{-4}}$
at low temperatures replacing it by the phenomenological law ${\rho(E)=\rho_{0}(E/E_{max})^{\mu}}$
with a small $\mu\approx0.3$. It is very likely that this law is
a consequence of a more complicated, than assumed usually, nature
of the TLS in physical glasses. The data also indicate that interaction
between TLS is responsible for their dephasing and the noise generated
by them. 

This work was supported by the NSA under ARO W911NF-13-1-0431, Templeton
foundation (40381) and ANR QuDec. We want to thank J. Burnett, T.
Lindström, S. T. Skacel, K. D. Osborn, J. Gao, J. Martinis, A. Ustinov, W. Oliver for many interesting discussions.
%

\end{document}